\documentclass [aps,twocolumn,showpacs]{revtex4-1}
\usepackage{graphicx}
\usepackage{textcomp}
\usepackage{psfrag}
\usepackage{epstopdf}
\usepackage{float}

\begin{document}
\title{General framework for acoustic emission during plastic deformation} 
\author{Jagadish Kumar$^1$, Ritupan Sarmah$^2$ and G. Ananthakrishna$^3$}
\affiliation{$^1$Department of Physics, Utkal University, Bhubaneswar 751004, India \\
 $^2$Department of Physics, Tezpur University, Tezpur-784028, India\\
 $^3$Materials Research Centre, Indian Institute of Science, Bangalore 560012, India}
\begin{abstract}
Despite the long history, so far there is no general theoretical framework for calculating the acoustic emission  spectrum accompanying any plastic deformation. We set up a discrete wave equation with plastic strain rate as a source term and include the Rayleigh-dissipation function to represent dissipation accompanying acoustic emission.  We devise  a method of bridging the widely separated time scales of plastic deformation and elastic degrees of freedom. While this  equation is applicable to any type of  plastic deformation, it  should   be supplemented by evolution equations for the dislocation microstructure for  calculating the plastic strain rate. The efficacy of the framework is illustrated by considering three distinct cases of  plastic deformation. The first one is  the acoustic emission during a typical continuous yield exhibiting a smooth stress-strain curve. We first construct an appropriate set of evolution equations for two types of dislocation densities and then show that the shape of the model stress-strain curve and accompanying acoustic emission spectrum match very well with  experimental results.  The second  and the third are  the more complex cases of the Portevin-Le Chatelier bands  and  the  L{\"u}ders band. These two cases are  dealt with in the context of the Ananthakrishna model  since the model predicts the three types of the Portevin-Le Chatelier bands  and also L{\"u}ders-like  bands. Our results show that for the type-C bands where the serration amplitude  is large, the acoustic emission spectrum consist of well separated bursts of acoustic emission. At higher strain rates of hopping type-B bands, the burst type acoustic emission  spectrum tends to overlap forming a nearly continuous background with some sharp acoustic emission  bursts. The latter can be identified with the nucleation of new bands. The acoustic emission spectrum associated with the continuously propagating type-A band is continuous.  These predictions are consistent with experimental results.  More importantly, our study  shows that the  low amplitude continuous acoustic emission spectrum seen in both the type-B and A band regimes is directly correlated to  small amplitude serrations induced by propagating bands. The acoustic emission spectrum of the L{\"u}ders-like band matches with recent experiments as well. In all of these cases, acoustic emission signals are burst-like  reflecting the intermittent character of dislocation mediated plastic flow.
\end{abstract}
\pacs{43.40.Le, 62.20.fq, 05.45.-a, 83.50.-v}
\maketitle
\section{Introduction} 
Two striking features of acoustic emission are its intermittent  character and its occurrence in a surprisingly large variety of systems ranging from geological scales to laboratory scales.  A good example from the geological scale is the acoustic emission (AE)  during volcanic activity  \cite{Diodati91}.  Varied  laboratory scale examples  such as AE from crack nucleation and propagation in fracture of solids \cite{Lockner96,Petri94,Rumiepl}, thermal cycling of martensites \cite{Planes,Rajeevprl,Kalaprl}, peeling of an adhesive tape \cite{MB,Cicc04,Rumiprl,Jag08} and  collective dislocation motion can be cited \cite{Migue1,Weiss,Jagprl}. Clearly, while the sources that lead to AE signals in such widely different situations are necessarily different,  they are generally attributed to the release of stored elastic energy in the system. Further, the AE spectrum in all of these cases is intermittent, a feature reflective of the underlying jerky motion of the sources generating the AE signals.  The phenomenon has been effectively used as a non-destructive tool in locating the sources and mechanisms generating the AE signals \cite{Ser03}. The method involves recording the arrival times of a wave at multiple transducers which in turn determines the distances of the AE source from the transducers. This procedure is  akin to that  adopted in fracture studies on rock samples \cite{Lockner96,Weiss}.  This method has been used to explain the   power law distribution  of the amplitudes of  the AE signals in the deformation studies of ice samples \cite{Migue1,Weiss}. 

Considerable insight into the intermittent character of dislocation mediated plastic deformation has come from acoustic emission measurements \cite{Ser03,Migue1,Weiss}. Indeed, such AE  studies carried out for over five decades  have  established specific  correlations between the nature of the AE signals and the stress-strain curves for  different situations \cite{Dunegan69,James71,Caceres87,Han11,Zeides90,Chmelik02,Chmelik07,Zuev08,Shibkov11}. However, there is lack of clarity as to why such distinct correlations exist \cite{Dunegan69,Han11,James71,Caceres87,Zeides90,Chmelik02,Chmelik07,Leby12}. For instance, even  early studies on the AE spectrum  for  the smooth homogeneous yield phenomena showed a intermittent AE spectrum  \cite{Dunegan69}. Improved techniques confirm the pulse like character of the AE events. The  general shape of the AE spectrum for this case exhibits a  peak just beyond the elastic regime decaying for larger strains \cite{Dunegan69,Han11}. Since the stress-strain ($\sigma-\epsilon$) curves remain smooth, the  pulse like acoustic emission signals are attributed to the intrinsic intermittent motion of dislocations.  Then, the smooth $\sigma-\epsilon$ curves are interpreted as resulting from the averaging process of the dislocation activity in the sample. Indeed, the  intermittent character of  dislocation motion at the microscopic level is reflected in the strong stress fluctuations seen in nanometer  sized samples that are  not seen in macroscopic samples \cite{Uchic09}.

 In contrast,   the nature of the AE spectrum is qualitatively different for the case of discontinuous flows where the stress-strain curves display stress-serrations.  For example, studies on the Portevin-Le Chatelier (PLC) effect, a kind of propagative instability, have established specific types of correlations  between the AE spectrum and the  different types of deformation bands and the associated stress-strain curves \cite{James71,Zeides90,Chmelik02,Chmelik07,Leby12}.   Similar correlations exist for the L{\"u}ders band \cite{Chmelik02,Chmelik07,Zuev08,Shibkov11}, another type of propagative instability \cite{Neuhausser83}. Furthermore, the AE spectrum for  the L{\"u}ders band \cite{Neuhausser83} is different from that for the three types of PLC bands \cite{GA07,Yilmaz11}. In these cases of propagative instabilities, collective dislocation processes  govern the nature of the bands and the associated stress-strain curves.   Thus, there is  a necessity to simultaneously describe the collective behavior of dislocations and  the wave equation that captures the inertial time scale. 

Early  theoretical attempts to explain AE  during plastic deformation were based on the AE response of individual dislocation mechanisms   such as  the Frank-Reed source \cite{Malen74,Tir92,Polyzos95}.   However, such methods are clearly unsuitable while following the AE signals  during the entire course of deformation since the AE sources themselves evolve as dislocations multiply and interact with each other. Clearly,  the AE spectrum from collective dislocation phenomenon such as the PLC effect and L{\"u}ders band cannot be explained  as a superposition of individual dislocation contributions.

The purpose of the present  paper is to set up a general mathematical framework for describing the acoustic emission for {\it any type} of plastic deformation.  Devising such a theory involves developing a method for dealing  with  widely separated time scales of plastic deformation and  the inertial time scale,  and a  method for describing collective effects of dislocations manifested in the PLC effect and L{\"u}ders bands \cite{GA07,Bhar02,Bhar03,Bhar03a,Anan04,Ritupan15}.   In  a preliminary short communication, we outlined a way of dealing with both dislocation dynamics and elastic degrees of freedom  specifically applicable  to the PLC instability \cite{Jagprl}.  Our present approach involves several  mathematical steps  such as (a) including  a dissipative term representing the acoustic energy,  (b) using the plastic strain rate  as a source term in the wave equation for the elastic degrees of freedom, (c) setting-up evolution equations for dislocation microstructure, and (d) imposing mutually compatible boundary conditions on both the wave equation  and the evolution equations for the dislocation microstructure. As  we shall show, point (d) requires describing the wave equation at a discrete level. 

Consider the  functional form  for the dissipated acoustic energy in terms of a relevant 'state variable' that  also evolves as  deformation proceeds. To be applicable to any plastic deformation, we require  the functional form for the dissipated AE energy  to be independent of the nature of the deformation process or the evolving microstructure, but such that it could   be coupled to the evolution equations for the dislocation microstructure. Indeed, an expression  for the dissipated acoustic energy  was introduced while modeling the power law  distribution of AE signals during thermal cycling of martensites  \cite{Rajeevprl,Kalaprl}. The idea was that, to a leading order, dissipated acoustic energy could be represented by the  Rayleigh dissipation function \cite{Land}. This choice  proved quite successful in explaining the AE spectrum in a number of situations including the power law distribution  of the AE signals during  martensite transformation \cite{Rajeevprl,Kalaprl}, peeling of an adhesive tape \cite{Rumiprl,Jag08} and crack propagation \cite{Rumiepl}. However, in these cases, only elastic or viscoelastic degrees of freedom having similar order time scales had to be described. In contrast,  plastic deformation is more complicated  since it requires describing widely separated time scales of plastic deformation and inertial time scale.

The efficacy of the framework is illustrated for  three cases. First  is the acoustic emission  during  a continuous yield, second is during  the PLC  bands and the third is during  the  L{\"u}ders band. For the first  case, we set-up a dislocation dynamical  model that uses two types of dislocation densities to predict the smooth stress-strain curve and also the general shape of the AE spectrum and its burst like character. For the second and third cases we  use  the Ananthakrishna (AK) model for the PLC effect \cite{Anan82} since it predicts the most generic features of the PLC effect including the three band types \cite{Anan82,Bhar02,Bhar03,Anan04,Ritupan15} and also L{\"u}ders-like bands \cite{Ritupan15}.   The AE bursts for the  uncorrelated type-C bands are well separated as the type-C stress drops.  For the hopping type-B bands, the AE bursts  overlap forming low amplitude nearly continuous background  AE signals. More importantly,  we find sharp bursts of AE  superposed on the  low level continuous AE background that can be unambiguously  identified with the nucleation of new bands. For the type-A propagating band, we find a continuous AE spectrum. All of these features are consistent with the experimental AE spectrum \cite{Zeides90,Chmelik02,Chmelik07}.  For the L{\"u}ders band,  the nature of the AE spectrum predicted is again consistent with recent experiments \cite{Chmelik02,Chmelik07,Zuev08}.
tin
\section{A general framework for acoustic emission during plastic deformation}

We begin by constructing a  wave equation that includes the contribution from dissipated acoustic energy. For the sake of simplicity, we work in one dimension. The physical mechanism attributed to  the generation of AE signals during plastic deformation can be broadly termed as 'dislocation multiplication mechanisms' such as the Frank-Reed source, the  abrupt unpinning of dislocations from pinning points or  from solute atmosphere as in the case of the PLC effect. These mechanisms set-off local elastic disturbances.  There are dissipative forces that tend to oppose the growth of the elastic disturbances so that mechanical equilibrium is restored.
We represent the dissipative energy  \cite{Rumiepl,Rajeevprl,Kalaprl,Rumiprl,Jag08} by the  Rayleigh dissipation function \cite{Land}   given by $ {\cal R}_{AE} = {\eta\over2}\int\Big[{\partial \dot\epsilon_e(y) \over \partial y} \Big]^2 dy $. Here,  $\eta$ is the damping co-efficient. Noting that  ${\cal R}_{AE} \propto \dot \epsilon_e^2(t)$, we interpret  $ {\cal R}_{AE}$ as the acoustic energy that is dissipated during the abrupt motion of  dislocations \cite{Rumiepl,Rajeevprl}. 

We now set up the wave equation for the elastic strain $\epsilon_e$ for a one dimensional crystal. The Lagrangian consists of the kinetic energy of the crystal $T= {\rho\over 2} \int\Big[{\partial \epsilon_e(y) \over \partial t} \Big]^2 dy$, with $\rho$ referring to the density of the material, the strain energy $ V_{loc}={\mu\over2} \int\Big[{\partial \epsilon_e(y) \over \partial y} \Big]^2 dy$ with $\mu$ referring to the elastic constant, and the gradient  energy  $V_{grad}=\frac{D}{4}\int\Big[{\partial^2\epsilon_e(y) \over \partial y^2} \Big]^2 dy$, where $D$ is the strain gradient coefficient. $V_{grad}$ makes the sound wave dispersive, a term that is  particularly important when localized transient waves are generated. This term may be regarded  as the next dominant term (to the strain energy) in the Ginzburg-Landau expansion of the free energy.
Then, using the Lagrangian ${\cal L}= T-V_{loc}-V_{grad}$ in the Lagrange equations of motion
\begin{eqnarray}
{d\over dt}\left({\delta{\cal L} \over {\delta\dot\epsilon_e(y) }}\right)-
{\delta{\cal L} \over {\delta\epsilon_e(y) }}+
{\delta{\cal R} \over {\delta\dot\epsilon_e(y)}}&=& 0.
\end{eqnarray}
we have,  
\begin{eqnarray}
\rho \frac{\partial^2 \epsilon_e}{\partial t^2} = \mu \frac{\partial^2 \epsilon_e}{\partial y^2} +\eta \frac{\partial^2 {\dot \epsilon_e}}{\partial y^2}- D \frac{\partial^4 \epsilon_e}{\partial y^4}.
\label{wave}
\end{eqnarray}
Equation (\ref{wave}) describes sound waves in the absence of dislocations. However, during plastic flow, transient elastic waves (or acoustic emission) are triggered by the abrupt motion of dislocations, which then propagate through the elastic medium. This can be described by including plastic strain rate as a source term in Eq. (\ref{wave}). Then,  the relevant (inhomogeneous) wave equation describing the acoustic emission process takes the form \begin{eqnarray}
\rho \frac{\partial^2 \epsilon_e}{\partial t^2} = \mu \frac{\partial^2 \epsilon_e}{\partial y^2} - \rho  \frac{\partial^2 \epsilon_p}{\partial t^2}+\eta \frac{\partial^2 {\dot \epsilon_e}}{\partial y^2}- D \frac{\partial^4 \epsilon_e}{\partial y^4}.
\label{comp_wave}
\end{eqnarray}
Here $c= \sqrt{\mu/\rho}$ is the velocity of sound and $\dot\epsilon_p(y,t)$ is the plastic strain rate. Note that $\dot\epsilon_p(y,t)$ is a function of both space and time and hence contains full information of any  heterogeneous character of the deformation (as for  the PLC effect),  which has to be  calculated by  setting-up appropriate evolution equations for suitable types of dislocation densities.

We note here that Eq. (\ref{comp_wave}) has the standard form of  a partial differential equation  with $\dot\epsilon_p$  acting as a source term. This form excludes the possibility of  transient acoustic waves (generated by  the source $\dot\epsilon_p$ itself)  influencing  the plastic strain rate or equivalently dislocations or collective  dislocation motion as the case may be. Such an effect is at best a second order effect.

Finally, we need to specify the initial and boundary conditions of Eq. (\ref{comp_wave}). First, the constant strain rate condition is imposed by fixing one end of the sample  and applying a  traction at the other end. Second, the boundary conditions  of  Eq. (\ref{comp_wave}) should be consistent with those imposed on the evolution  equations for the dislocation densities, in which the latter is determined by physically meaningful values for the dislocation densities. Then,  $\dot\epsilon_p(y,t)$  obtained near the boundary sites need not be  consistent with those  imposed on  Eq. (\ref{comp_wave}).  Third,  the machine stiffness gripping the ends of the sample is higher than that of the sample, which is not easy to include in  Eq. (\ref{comp_wave}). In fact, conventionally,  the information about the machine stiffness (for example, in the constant strain rate case) goes only in the effective modulus of the machine and the sample.  Therefore,  we start with a Lagrangian defined on a grid of $N$ points and derive a discrete set of wave equations. This method allows us to make a distinction between points well within the sample and those at the boundary where the machine grips the sample. The method also brings clarity to the boundary conditions. 
\label{Dis-wave-eqn}
\begin{figure}[!h]
\centerline{\includegraphics[height=1.50cm,width=8.5cm]{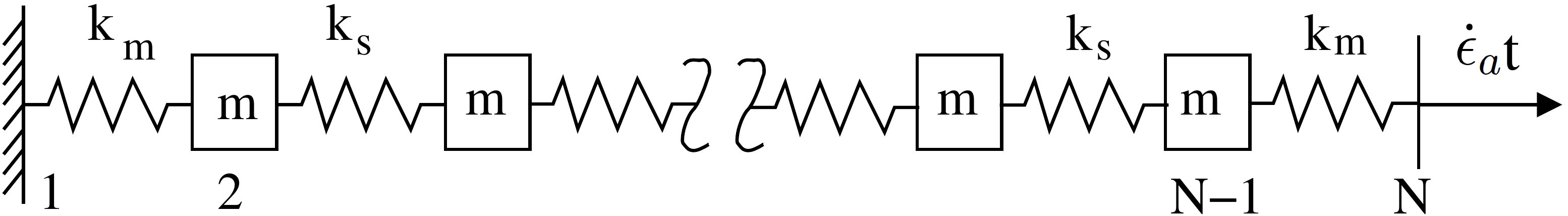}} 
\caption{Mechanical model for the specimen fixed at one end and pulled at a constant strain rate at the other end.}
\label{mass_spring} 
\end{figure} 
\subsection{Discrete form of the wave equation}

Consider a sample of length $L$  deformed in a constant strain rate condition schematically  represented by a spring-block system of $N$ points of mass $m$ coupled to each other through  spring constant $k_s$ as shown in Fig. \ref{mass_spring}. Then, the condition that the sample is gripped at the ends translates into using a different spring constant $k_m$ for the end springs.  Let $a$ be the separation between the points in the undistorted state. Then, the local displacements from the equilibrium  positions are the dynamical variables of interest.  However, since we use plastic strain rate for describing plastic deformation, we use strain variables. Therefore, we define the  strain variables $\epsilon_e(i)$ and their time derivative ${\dot\epsilon}_e(i)$ for each of these points. Further,  the condition that the sample is pulled at a constant strain rate is imposed by fixing the first point and pulling the $N^{th}$ point at a constant strain rate $\dot\epsilon_a$. Then, we can define a Lagrangian for the system of $N$ points. The kinetic energy $T$  of the system is 
\begin{eqnarray}
T=\sum_{i=2}^{N-1} \frac{1}{2}m{ \dot{\epsilon}_e^2(i)}.
\end{eqnarray}
Here the over dot refers to the time derivative. 
The local potential energies $V_{loc}$ and $V_{grad}$ are respectively given by 
\begin{eqnarray}
\nonumber
V_{loc}=\sum_{i=2}^{N-2} \frac{k_s}{2}\Big[{\epsilon_e(i+1)- \epsilon_e(i)}\Big]^2 + \frac{k_m}{2} \epsilon_e^2(2)\\
+ \frac{k_m}{2}\Big[ \epsilon_e(N-1) - \epsilon_e(N) \Big]^2.
\end{eqnarray}
\begin{eqnarray}
\nonumber
V_{grad}&=&\sum_{i=3}^{N-2} \frac{D}{2}  \Big[ \epsilon_e(i+1)+\epsilon_e(i-1)-2\epsilon_e(i) \Big]^2\\
\nonumber
&+&  \frac{D}{2} \Big[ \epsilon_e(3)-2\epsilon_e(2) \Big]^2 + \frac{D}{2} \epsilon_e^2(2) +  \frac{D}{2}\epsilon_e^2(N-1). \\
&+&  \frac{D}{2}\Big[ \epsilon_e(N)-2\epsilon_e(N-1) + \epsilon_e(N-2) \Big]^2. 
\end{eqnarray}
The dissipated acoustic energy  is given by 
\begin{eqnarray}
\nonumber
 R_{AE}&=&\frac{\eta}{2}\sum_{i=2}^{N-2}\big(\dot{\epsilon}_e(i+1)- \dot{\epsilon}_e(i)\big)^2 
+ \frac{\eta}{2}\big[\dot{\epsilon}_e^2(2)\\
&+& \dot{\epsilon}_e^2(N-1)\big].
\label{Diss-finite}
\end{eqnarray}
Then, using the  Lagrange equations of motion we get
\begin{eqnarray}
\label{waveqn_discret1}
& &\ddot \epsilon_e(1) = 0.0,\\
\nonumber
\label{waveqn_discret2} 
& &\ddot \epsilon_e(2) = -\frac{c^2}{a^2} \big[ \{\epsilon_e(2)-\epsilon_e(3)\}
 +\frac{k_m}{k_s}\epsilon_e(2)\big] -\frac{\partial\dot{\epsilon}_p(2,t)}{\partial t} \\
\label{waveqn_discret3} 
& - & \frac{\eta'}{a^2\rho}\big[\dot \epsilon_e(2)-\dot \epsilon_e(3)\big]
+ \frac{D'}{a^4\rho}\big [\epsilon_e(4)+\epsilon_e(2)-2 \epsilon_e(3)\big], \\
\nonumber
& &\ddot \epsilon_e(3)= \frac{c^2}{a^2}\big[ \epsilon_e(4)+\epsilon_e(2)-2\epsilon_e(3)\big]-\frac{\partial\dot{\epsilon}_p(3,t)}{\partial t}\\
\nonumber
&+& \frac{\eta'}{a^2\rho} \{ \dot \epsilon_e(4)+\dot \epsilon_e(2)-2\dot \epsilon_e(3)\}\\
&-& \frac{D'}{a^4\rho} \{\epsilon_e(5)-4\epsilon_e(4)+ 5 \epsilon_e(3)- 2 \epsilon_e(2) \},\\
\nonumber
\label{waveqn_discret}
& &\ddot \epsilon_e(i)= \frac{c^2}{a^2} \{\epsilon_e(i+1)-2 \epsilon_e(i)+ \epsilon_e(i-1)\}- \frac{\partial\dot{\epsilon}_p(i,t)}{\partial t}\\
\nonumber
&+&  \frac{\eta'}{a^2\rho} \{ \dot \epsilon_e(i+1)-2\dot \epsilon_e(i) + \dot \epsilon_e(i-1)\}- \frac{D'} {a^4\rho}\big [\epsilon_e(i+2)\\
&-& 4 \epsilon_e(i+1)+ 6 \epsilon_e(i) - 4 \epsilon_e(i-1) + \epsilon_e(i-2)\big ],\\
\label{waveqn_discret99}
\nonumber
& &\ddot \epsilon_e(N-1)= - \frac{c^2}{a^2}\big[ \{\epsilon_e(N-1)-\epsilon_e(N-2)\}\\
\nonumber
&-& \frac{k_m}{k_s} \{\epsilon_e(N)-\epsilon_e(N-1)\}\big] -\frac{\partial\dot{\epsilon}_p(N-1,t)}{\partial t}\\
\nonumber
&+ &\frac{\eta'}{a^2\rho} \big[\dot \epsilon_e(N)+ \dot \epsilon_e(N-2)-2\dot \epsilon_e(N-1)\big] -\frac{D'}{a^4\rho} \big[\epsilon_e(N-3)\\
&-& 4\epsilon_e(N-2)+ 5 \epsilon_e(N-1)-2 \epsilon_e(N) \big].\\
\nonumber
\end{eqnarray}
Equation (\ref{waveqn_discret}) is valid for $i= 4$ to $N-1$.  Using $\rho = m/a^3$ and  appropriate length factors of  $a$ we retain the definition of $c^2 = \mu/\rho$ (see Eq. (\ref{comp_wave})) with $\mu=k_s/a, \eta'=\eta/a$ and $D'=Da$. Note that  $\dot\epsilon_p$  has been included as a source term in Eqs. (\ref{waveqn_discret1}-\ref{waveqn_discret99}). Equations  (\ref{waveqn_discret1}-\ref{waveqn_discret99}) are  solved on a grid of  $N$ points with appropriate initial and boundary conditions. The initial conditions are
\begin{equation}
\nonumber
\epsilon_e(1,0)=  0 ;   \epsilon_e(i,0) = 0 + \xi \times \epsilon_r, \quad i=2,..,N-1, 
\label{IC1} 
\end{equation}
where $\epsilon_r$ represents the strain due to inherent defects in the sample and $ \xi$ is a random number in the interval  $- \frac{1}{2} < \xi < \frac{1}{2}$. Here, we use $\epsilon_r \sim 10^{-7}$. The boundary condition that the left end is fixed and the right end is being pulled at a constant strain rate $\dot\epsilon_a$ can be written as 
\begin{eqnarray}
\epsilon_e(1,t) = 0, \quad \epsilon_e(N,t) = \dot{\epsilon}_a t; 
\,\, t>0,
\label{BC1}
\end{eqnarray}

\section{Dislocation dynamical models for plastic deformation}

These steps are clearly applicable to any plastic deformation situation, but  must be supplemented by constructing dislocation based models to calculate the source term $\dot\epsilon_p$  in the wave equation.  The types  of dislocation densities that must be used and the nature of the equations  are dictated by the kind of plastic deformation considered, namely the continuous yield and the two propagative instabilities \cite{GA07,Anan82,Bhar02,Bhar03,Anan04,Ritupan15}. The dynamical approach followed here has the ability to use experimental $\dot\epsilon_a$ unlike in simulations where the imposed strain rates are several orders of magnitude higher \cite{Ritupan15}. Further, we can also adopt other experimental parameters used in experiments, for instance $\sigma_y, E^*$ etc. 

\subsection{A dislocation dynamical model for a continuous yield point phenomenon} 
\label{dd-model-c-yield}

We first construct a  model that uses  two types of dislocation densities, namely, the mobile $\rho_m$ and the immobile (or the forest)  density $\rho_{im}$  that reproduces a typical smooth stress-strain curve. Most  dislocation mechanisms used are drawn from the AK model for the PLC effect \cite{Anan82,GA07,Bhar02,Bhar03,Bhar03a,Anan04,Ritupan15}.  They  can be broadly categorized into dislocation multiplication and  transformation processes.   As dislocations multiply (due to the double cross-slip process), they interact with each other to form  dipoles and junctions \cite{Kubin90}. They can also annihilate. Each of these mechanisms act as a growth or loss processes for $\rho_m$ and $\rho_{im}$. The general form of multiplication of dislocations can be  written as $\theta V_m(\sigma_a) \rho_m$ with $V_m(\sigma_a)$ representing the mean velocity of  dislocations. $\theta$ is the inverse of a length scale that physically represents points from which the line length of dislocations multiplies (see Ref. \cite{GA07} and \cite{Ritupan15} for details).  Several phenomenological expressions have been suggested for $V_m(\sigma_a)$ \cite{Neuhausser83}. Here, we use  $V_m(\sigma_a) = v_0 \big[\frac{\sigma_{eff}}{\sigma_m}\big]^m$,  where $\sigma_{eff}= \sigma_a - h \rho_{im}^{1/2}$. Here, $m$ is a velocity exponent, $h \rho_{im}^{1/2}$ the back stress. The parameter  $h = \alpha G b$ where $\alpha\sim 0.3$ is a constant, $b$ the magnitude of the Burgers vector and  $G$  the shear modulus. Indeed, one can  rewrite the multiplication rate as $\nu_m \rho_m = \nu_0  \big[\frac{\sigma_{eff}}{\sigma_m}\big]^m \rho_m$, where $\nu_0= \theta v_0$. The formation of  dipoles  occurs when two dislocations moving in nearby glide planes approach a minimum distance (typically a few nanometers) acts as a loss term to $\rho_m$. This is represented by $\beta \rho_m^2$, where $\beta$ has dimensions of the rate of area swept-out by dislocations. Similarly, the   annihilation of a mobile  dislocation with an immobile one is represented by the term $f\beta \rho_m \rho_{im}$ with a rate $f\beta$, where $f$ is a  dimensionless parameter. This term is generally small compared to other loss terms for $\rho_m$ and therefore $f << 1$.  Finally, dislocations moving in different glide planes intersect each other to form junctions. This is a loss term to $\rho_m$ given by   $ \Theta \rho_m  \rho_{im} $. Here,    $\Theta$ is a parameter that however, depends on the mean separation between junctions themselves that also  evolves as deformation proceeds (i.e., $\rho_{im}$ increases). Then, $\Theta \propto 1/\rho_{im}^{1/2}$ or $\Theta = \delta\rho_{im}^{-1/2}$.  Here $\delta$ is considered constant since main contribution to $\Theta$ has been absorbed.  Then, the loss term for  $\rho_m$ is  $ \delta\rho_m \rho_{im}^{1/2}$. ($\delta$  has the dimension of velocity.) This represents the forest mechanism \cite{Kubin90}. This acts  as a gain term to $\rho_{im}$.  Then the evolution equations are
\begin{eqnarray}
\label{NM}
\nonumber
\frac{\partial\rho_m}{\partial t'} 
& = & \theta v_0 \rho_m [\frac{1}{\sigma_y}(\sigma_a - h \rho_{im}^{1/2})]^m -\beta \rho_m^2 - f \beta \rho_m \rho_{im}\\
 &-&\delta \rho_m \rho_{im}^{1/2} + \frac{\Gamma\theta v_0}{\rho_{im}} \frac{\partial^2 }{\partial x^2}\big[\frac{\sigma_{eff}}{\sigma_y}\big]^m  \rho_m\\
\label{NIM}
\frac{\partial\rho_{im}}{\partial t'}  &=& \beta \rho_{m}^2 - f \beta \rho_m \rho_{im} + \delta \rho_m \rho_{im}^{1/2}.
\end{eqnarray}
(We use the primed time variable for  plastic strain rate calculations.) The spatial coupling in Eq. (\ref{NM}) arises since cross-slip allows dislocations to spread into neighboring regions. The factor $1/\rho_{im}$ prevents dislocations from  moving  into regions of high dislocation density \cite{Anan04}. 

These equations are coupled to the machine equation \cite{Hull} that enforces the constant strain rate  condition 
\begin{equation}
\frac{d\sigma_a}{dt'} = E^* \big[{\dot\epsilon}_a -\frac{b}{L}\int_0^L  v_0 \big[\frac{\sigma_{eff}}{\sigma_y}\big]^m \rho_m dx\big] = E^*[{\dot\epsilon}_a - \dot\epsilon_p(t')].
\label{S-eqn}
\end{equation}
\subsection{The Portevin-Le Chatelier effect and Ananthakrishna model } 
\label{AK-model}

We first summarize relevant features of the PLC effect \cite{GA07,Yilmaz11}. The PLC instability is seen in a window of strain rates and temperatures when samples of dilute metallic alloys are deformed under constant strain rate conditions.  It is characterized by three types of  bands and the associated stress-serrations \cite{GA07,Yilmaz11,Chihab87,Ranc05,Ranc08,Jiang05,Jiang07} observed with increasing strain rate   or decreasing temperature.  At the lower end of  $\dot\epsilon_a$, randomly nucleated static type-C bands with large characteristic stress drops are seen. The serrations  are quite regular. At intermediate $\dot\epsilon_a$,  'hopping' type-B bands are seen where a new band is formed ahead of the previous one giving a visual impression of a hopping character. The serrations are more irregular with amplitudes that are smaller than that for type-C.   Finally, at high $\dot\epsilon_a$,  the continuously propagating type-A bands associated with small stress drops are found.  These bands have been shown to  represent different  correlated states of dislocations in the bands \cite{Bhar02,Bhar03,Bhar03a,Anan04,Ritupan15}. 

There are number of models that target specific features of the PLC effect. These use  local strains, strain rates, negative SRS of the flow stress, activation enthalpy of dynamic strain aging, waiting etc \cite{Penning72,McCormic88,Kocks81,Kubin90,Hahner02,Kok03}. However, there are fewer models that predict the characteristic features of the three types of bands   \cite{Bhar03,Bhar03a,Anan04,Ritupan15,Hahner02,Kok03} required for calculating  the associated AE spectra. Here, we use  the AK model since it captures most  generic features of the PLC effect including the three types of bands and large number of features such as the existence of the instability within a window of strain rates, the negative strain rate behavior of the flow stress \cite{Anan82,Rajesh}, chaotic nature of stress drops at low strain rates \cite{Anan83} and the power law distribution of stress drop magnitudes and durations \cite{Noro97,Anan99,Bhar01,Bhar02,Bhar03,Anan04}. In addition, the AK model been recently shown to predict L{\"u}ders-like band as well \cite{Anan82,Bhar02,Bhar03,Anan04,Ritupan15}. The basic idea of the model is that all the qualitative features of the PLC effect emerge from the nonlinear interaction of  a few collective degrees of freedom  assumed to be represented by a few dislocation densities. The model consists of three types of densities, namely the mobile, immobile, and dislocations with solute atoms denoted by $\rho_m(x,t)$, $\rho_{im}(x,t)$ and $\rho_c(x,t)$ respectively.   The  evolution equations for these densities in the unscaled form are 
\begin{eqnarray}
\nonumber
\label{X-eqn}
\frac{\partial \rho_m}{\partial t'} &=& -\beta \rho_m^2 - f\beta \rho_m\rho_{im} -\alpha_m \rho_m + \gamma\rho_{im}  \\ 
&+ &\theta v_0 \big[\frac{\sigma_{eff}}{\sigma_y}\big]^m \rho_m + \frac{\Gamma\theta v_0}{\rho_{im}} \frac{\partial^2 }{\partial x^2}\big[\frac{\sigma_{eff}}{\sigma_y}\big]^m  \rho_m, \\
\label{Y-eqn}
\frac{\partial \rho_{im}}{\partial t'} &=& \beta \rho_m^2 -p \beta \rho_m\rho_{im} -\gamma\rho_{im} + \alpha_c\rho_c,\\
\label{Z-eqn}
\frac{\partial \rho_c}{\partial t'} &= & \alpha_m\rho_m - \alpha_c\rho_c 
\end{eqnarray}
All terms in Eq. (\ref{X-eqn}) except the third and fourth terms have been already explained (see section \ref{dd-model-c-yield}).    The third term $\alpha_m\rho_m$ in Eq. (\ref{X-eqn}) refers to solutes diffusing to  mobile dislocations temporarily arrested by immobile (forest) dislocations. 
Thus,  $\alpha_m \rho_m$  is the gain term for $\rho_c$. $\alpha_m$  is a function of the solute concentration  $C$ at the core of dislocations, $D_c$ the diffusion constant of the solute atoms and $\lambda$ an effective attractive distance  for the solute segregation. Then, $\alpha_m= \frac{D_c (T) C}{\lambda^2}$.   As dislocations progressively acquire more solute atoms, they slow down at a rate $\alpha_c$ and eventually stop at which point they are considered as $\rho_{im}$. Thus, the loss rate  $\alpha_c \rho_c$ in Eq. (\ref{Z-eqn})  is the gain term Eq. (\ref{Y-eqn}) for  $\rho_{im}$.  {\it  For the same reason, we consider  $\rho_{im}$ to include dislocations pinned by solute atmosphere as well.} (Note the difference in the interpretation of $\rho_{im}$ used in Eq. (\ref{NIM}) and Eq.  (\ref{Y-eqn}).) Thus, the loss term $\gamma \rho_{im}$ in Eq. (\ref{Y-eqn}) is a gain term in Eq. (\ref{X-eqn}). This term  is considered to represent the unpinning of that fraction of  immobile dislocations from the  solute clouds. As in the model for continuous yield (section \ref{dd-model-c-yield}), the spatial coupling (the sixth term in Eq. (\ref{X-eqn})) in this model arises from double cross-slip process that allows dislocations to move into neighboring spatial elements.  Equations (\ref{X-eqn}-\ref{Z-eqn}) are coupled to  the machine equation Eq. (\ref{S-eqn})  that represents the constant strain rate deformation  condition. 

\section{Computing acoustic emission spectrum during  plastic deformation}

Now we consider the basic difficulty in describing  slow plastic deformation and fast sound wave propagation simultaneously. Experimental strain rates  are in the range $10^{-6}-10^{-2}/s$ while experimental AE frequencies are from KHz to MHz that differ by almost $10^8-10^{10}$. This difference translates into the difference in the time steps for integration of the dislocation density equations (or $\dot\epsilon_p$) and the wave equations (Eqs. (\ref{waveqn_discret1}-\ref{waveqn_discret99})).  For the sake of clarity, we use primed variable $t'$ for the dislocation density evolution equations or for plastic strain rate $\dot\epsilon_p(k,t')$.  Denoting the $i^{th}$ integration time step of  $\dot\epsilon_p(k,t')$ (with $k$ referring  to spatial coordinate) by  $\delta t'_i$, for the time interval between $t'_{i+1} < t' < t'_i$, we need to ensure that  $ \Lambda \delta t = \delta t'_i$ where $\delta t$ is the step size used for Eqs. (\ref{waveqn_discret1}-\ref{waveqn_discret99}) and $\Lambda >>1$. Then, we should  impose  $\frac{\partial\dot{\epsilon}_p(k,t)}{\partial t} = \Lambda^2 \frac{\partial\dot{\epsilon}_p(k,t')}{\partial t'}$.   $\Lambda$ would  be different for each type of the plastic deformation cases considered since the time step  for integration for dislocation density evolution equations depends on whether they are stiff or not. However, for our purposes, it would be adequate to use the mean value of  $\Lambda $. As for the spatial  part, the wave equations and the dislocation density evolution  equations are solved on the same spatial grid of 100  points (for $L=0.05m$). 
Further, since $\dot\epsilon_p(k,t')$ is calculated at much coarser time steps compared to Eqs. (\ref{waveqn_discret1}-\ref{waveqn_discret99}),  we need to use interpolated values for  $\dot \epsilon_p(k,t')$ in the source term.  We now outline the steps used for obtaining  the AE spectrum. 

Step 1: Solve Eqs. (\ref{NM}-\ref{S-eqn}) (or Eqs. (\ref{X-eqn}-\ref{Z-eqn} and \ref{S-eqn}) for the AK model)  for the entire  time interval and obtain ${\dot\epsilon}_p(k,t_i')$ and $\sigma_a(t'_i)$ using fixed or variable time step $\delta t'$ (as the case may be)   for $i,...,M$ and $k=1,..,100$. \\
Step 2: Start with $t=0$ along with the stated initial and boundary conditions and solve Eqs. (\ref{waveqn_discret1}-\ref{waveqn_discret99}) for the interval $0 < t <t'_i$ corresponding to integration time step $\delta t'$ in Eqs. (\ref{NM}-\ref{S-eqn})  (or Eqs. (\ref{X-eqn}-\ref{Z-eqn} and \ref{S-eqn}) for the AK model)  for that interval. This gives  $\epsilon_e(k, t)$ for $0< t< t'_i$. Repeat integration for successive time steps. \\
Step 3: The stress $\sigma_e(t)$ obtained using $\epsilon_e(k, t)$  (and using the elastic modulus $E$ for sample) would in principle be different from  that  obtained from the machine equation Eq. (\ref{S-eqn}), particularly in plastic regime.  Note also that we need to use the interpolated values of ${\dot\epsilon_p}(k,t'_i)$ as an input into Eqs. (\ref{waveqn_discret1}-\ref{waveqn_discret99}).\\
Step 4: The  dissipated acoustic energy  is calculated using Eq. (\ref{Diss-finite}). We note here that value of $\eta$ is not known. However, from Eq. (\ref{wave}) (or Eq. \ref{comp_wave}), we see $\eta/\rho \sim 1$, since $\eta/\rho >> 1$  corresponds to the over damped situation and $\eta/\rho << 1$ corresponds to the under damped case. We have used a fixed value of $\eta$ for the three cases so that the AE spectrum reflects the relative magnitudes.   

We  stress that the method for calculating the AE spectrum is  approximate since Eq. (\ref{S-eqn}) assumes equilibration  to obtain plastic strain rate  ${\dot\epsilon}_p(k,t')$. The method is akin to adiabatic schemes.

\section{Acoustic emission spectrum during  a typical yield}

Computation of AE spectrum requires that we solve Eqs. (\ref{NM}-\ref{S-eqn}) and use the plastic strain rate ${\dot\epsilon}_p(k,t')$ as source term in Eqs. (\ref{waveqn_discret1}-\ref{waveqn_discret99}).  Therefore,  we first solve Eqs. (\ref{NM}-\ref{S-eqn}) for the entire  time interval to obtain ${\dot\epsilon}_p(k,t')$.

\subsection{Numerical solution of model equations for continuous yield}

We first estimate the orders of magnitudes of the model parameters. The parameters values of $E^*,\sigma_y,  b$ and $\dot\epsilon_a$ are taken from the targeted experiment. Here, we attempt to predict the smooth $\sigma-\epsilon$ curve in Fig. $2$c of Ref. \cite{Han11} for which experimental parameters are $\sigma_y=0.3GPa, E^*/\sigma_y=416$. (The value of  $b=0.25 nm$.) Experimental strain rate  $\dot\epsilon_a = 1.67 \times 10^{-4}/s$.  Theoretical parameter $\theta v_0$ constitutes a time scale which has been set to unity (one second) so that  the plastic strain rate evolution time scale matches the experimental time scale. Other  parameters $\beta,f,\delta$ are easily estimated by using  typical asymptotic values of $\rho_m$ and $\rho_{im}$ that are in the range $\sim 10^{13}-10^{14}/m^2$ \cite{Ritupan15}.  The results presented here are for parameter values given in the Table \ref{T1}.  The constant $\Gamma$ is of the order of $1/\beta $ and hence  $\Gamma\sim 10^{12}$.
\begin{table}[!h]
\caption{Parameter values used for the continuous yield model. }
\label{T1}
\begin{center}
\begin{tabular} {|c |c |c |c |c |c|}
\hline 
$ \beta(m^2s^{-1}) $ & $\delta(m s^{-1}) $ & $f$ & $ v_0 (ms^{-1}) $ & $ m$\\
\hline
$8.33 \times 10^{-14}$ & $5.4 \times 10^{-11}$ & $ 10^{-3}$ & $  10^{-7} $ & $10$\\ [0.1ex]
\hline
\end{tabular}
\end{center}
\end{table} 
Equations (\ref{NM}-\ref{S-eqn}) are solved on a grid of $N=100$ points for a sample length $L=0.05m$. The initial condition used for $\rho_m(j,0)$ and $\rho_{im}(j,0)$ are taken to be uniformly distributed along the sample with a Gaussian spread of their values around a mean value $4.5\times 10^{11}/m^2$ and $10^{12}/m^2$ respectively. The variance for $\rho_m$ is $1\times 10^{12}/m^2$ and that for $\rho_{im}$ is $1\times 10^{11}/m^2$. The boundary conditions are $\rho_m(1,t')=\rho_m(N,t')\sim  10^{11}/m^2$ and $\rho_{im}(1,t')=\rho_{im}(N,t') = 10^{14}/m^2$. The high value for $\rho_{im}(N,t')$  represents the fact that the sample is strained at the grips.  We have used 'ode15s' MATLAB solver  for the numerical solution. We shall use $\dot\epsilon_p(j,t')$  so obtained as  a source term in Eqs. (\ref{waveqn_discret1}-\ref{waveqn_discret99}) for the AE studies. However, since the step size needed for integrating  Eqs. (\ref{waveqn_discret1}-\ref{waveqn_discret99})  is $\delta t \sim 0.001$, it  requires that we supply the values of $\dot\epsilon_p(t')$ for intermediate times.  

The calculated model stress-strain curve  is shown in Fig. \ref{AE-Typical-Yield} along with experimental points extracted from Fig. 2(c) of \cite{Han11}. It is clear that the model $\sigma-\epsilon$ curve matches the experimental $\sigma-\epsilon$ quite well. 

\subsection{Acoustic emission spectrum}

\begin{figure}[!h]
\vbox{
\includegraphics[height=4.75cm,width=8.75cm]{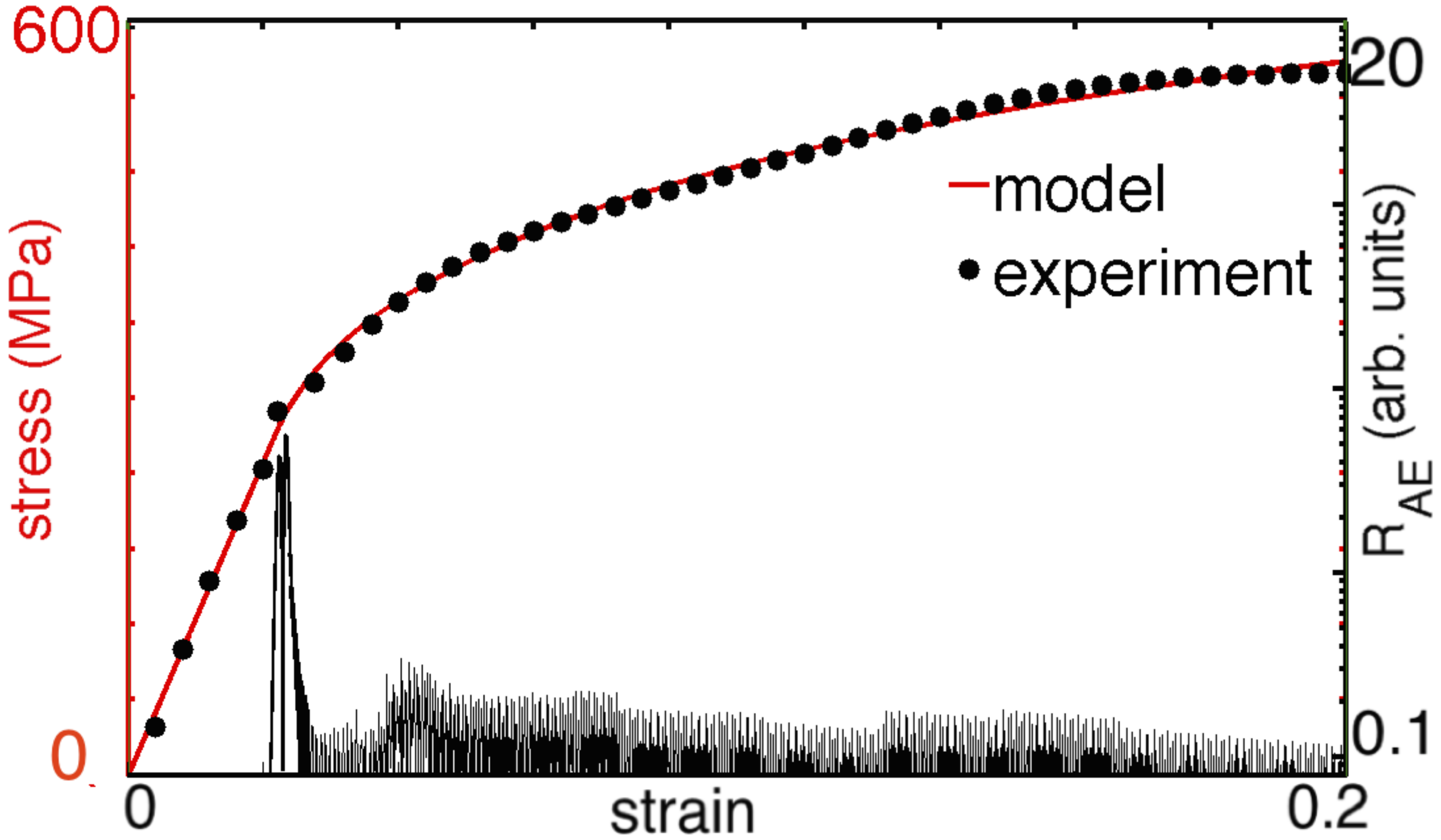}
}
\caption{(color online) Stress-strain curve for a continuous yield (continuous curve) with experimental points ($\bullet$) extracted from Fig. 2(c) of Ref. \cite{Han11} for $\dot\epsilon_a=1.67\times 10^{-4}/s$ along with the corresponding AE spectrum
}
\label{AE-Typical-Yield}
\end{figure}

We have used  $\dot\epsilon_p$ obtained from  Eqs. (\ref{NM},\ref{NIM}) and (\ref{S-eqn}) as a source term in Eqs.  (\ref{waveqn_discret1}-\ref{waveqn_discret99}) to obtain   the AE  spectrum $R_{AE}$ by using Eq. (\ref{Diss-finite}). This   is shown in Fig. \ref{AE-Typical-Yield}.   This may be compared with the experimental AE spectrum for the continuous yield shown in Fig. 2(c) of Ref. \cite{Han11}. It is clear that the overall shape of the model  AE spectrum is quite similar to  the experimental AE spectrum. Note that  even though  the $\sigma-\epsilon$ is smooth, the burst like character of the predicted AE signals (as also the experimental AE signals) are reflective of   the fundamentally intermittent nature of plastic deformation. 
\begin{table}[!h]
\caption{Parameter values used for the AK model. }
\label{T2}
\begin{center}
\begin{tabular} {|c |c |c |c |c |}
\hline 
$E^* (GPa) $ & $ \sigma_y(GPa) $ & $\alpha_m(s^{-1}) $ & $\alpha_c(s^{-1}) $ & $ v_0 (ms^{-1}) $ \\
\hline
$48$  & $0.2$ & $0.8$ & $0.08$ & $ 10^{-7} $  \\ [0.1ex]
\hline
$\gamma (s^{-1})$ & $  f$ & $m $ & $ \beta (m^{2}s^{-1}) $ & $\Gamma$ \\
\hline
$5\times 10^{-4}$  & $1$ & $3$ & $5 \times 10^{-14}$ & $10^{12}$ \\ [0.5ex]
\hline
\end{tabular}
\end{center}
\end{table} 
\section{Acoustic emission accompanying the different types PLC bands}

We first consider the solution of Eqs. (\ref{X-eqn}-\ref{Z-eqn}) and Eq. (\ref{S-eqn}), and discuss the  features of different type of PLC bands and the associated serrations predicted by the AK model before computing the acoustic energy  $R_{AE}$. 
To do this, we first  estimate the parameters following the same procedure adopted for the earlier case (section \ref{dd-model-c-yield}).  Experimental parameters such as  ${\dot\epsilon}_a, E^*, b$  and $h$   are adopted  from experiments. As for the theoretical parameters, the model time scale $\theta v_0$ is set  to unity as for the  previous case.  The other  parameters $ f\beta,\gamma,\alpha_m$ and  $\alpha_c$ are fixed easily by  providing the steady state values of $\rho_m,\rho_{im}$ and $\rho_c$ \cite{Ritupan15}. (Note that steady state exists for these set of equations.) As for $\alpha_m$, it is estimated by using $\alpha_m= \frac{D_c C}{\lambda^2}$.  The exact values used are shown in Table \ref{T2}.  

We solve Eqs. (\ref{X-eqn}-\ref{Z-eqn}) and  (\ref{S-eqn})  by using an adaptive  step size algorithm (ode15s MATLAB solver). The initial values of the dislocation densities are chosen much the same way as for the previous case. At the boundary, we use  two orders higher values for $\rho_{im}(j,t')$ at $j = 1$ and $N$ than the rest of the sample. Further, as bands cannot propagate into the grips, we use $\rho_m(j,t') =\rho_c(j,t')=0$ at $j=1$  and $N$. 

\subsubsection{The Portevin-Le Chatelier bands in the Ananthakrishna model}

Earlier studies have demonstrated that  the AK  model predicts the three types of bands C, B and  A with increasing strain rate \cite{Bhar02,Bhar03,Anan04,Ritupan15}. At low strain rates, uncorrelated static type-C bands are seen.  The corresponding stress-time plot displays  large amplitude nearly regular serrations as in experiments \cite{GA07,Yilmaz11,Chihab87,Ranc05,Ranc08,Jiang05,Jiang07}. 

As $\dot\epsilon_a$ is increased we see  hopping type-B bands.   The serrations are  irregular and  are of smaller  magnitude. One important  feature predicted by the model relevant to the AE studies  is the correlation between  band propagation property and  the small amplitude serrations (SAS's). In a recent study \cite{Ritupan15}, it was demonstrated that {\it band propagation induces small amplitude serrations that are bounded on both sides by large amplitude stress drops.}  Figure \ref{PLCAE_BAND_B}(a) shows  two type-B bands marked AB and CD.  The corresponding SAS's  induced by these two propagating bands are shown by the two sets of arrows AB and CD.  This stretch of SAS's are bounded on  both sides by large amplitude stress drops. The  large stress drop at A is well correlated  with the nucleation of the band AB. The one at B corresponds to stopping of the band. Similar observation hold for the band CD as well.   

As we increase $\dot\epsilon_a$, the extent of propagation increases. Concomitantly, the duration of small amplitude serrations increases.  Typical contour plots of two type-A bands marked ABC and DEF  are shown  in Fig. \ref{PLCAE_BAND_A}(a) for $\dot\epsilon_a=5.5\times 10^{-5}/s$.  The corresponding stretches of SAS's induced by the propagating bands are marked by three sets of arrows marked ABC and  DEF.  While the spatial 'width' of the propagating band is nearly uniform,  occasionally,  one  finds perturbations in the width. Two such  points B and E are shown. At these points we find  relatively large amplitude stress drops. Another feature is that the mean stress level of these SAS's increases or decreases as is clear for serrations A to B and C to D in  Fig. \ref{PLCAE_BAND_A}(a).  This feature  is seen in many experimental $\sigma-\epsilon$ curves at high strain rates. (See Fig. 1(c) for Cu-Al in Ref. \cite{Anan99}.) As we shall see, these features have a direct influence on the nature of AE spectrum.
\begin{figure}[!h]
\vbox{
\includegraphics[height=4.0cm,width=8.4cm]{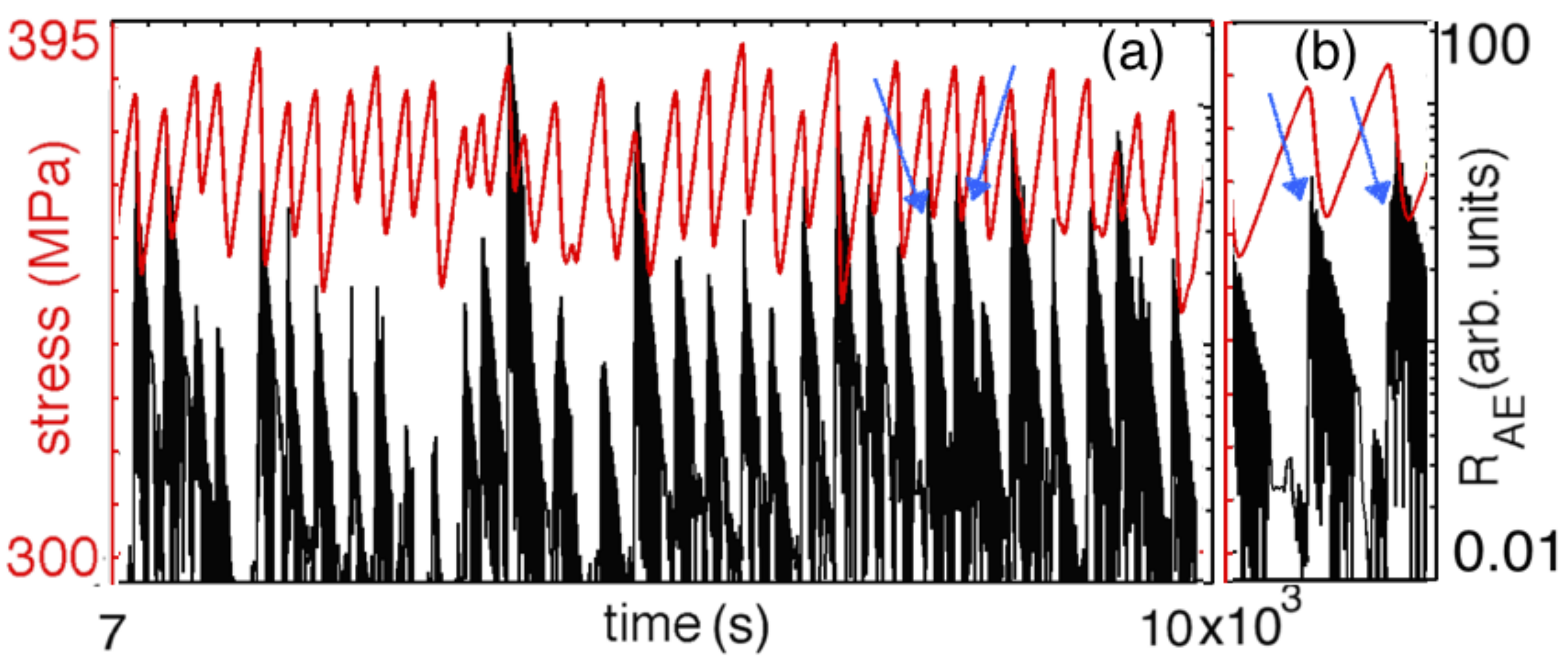}
}
\caption{(color online)  (a)  Stress-strain curve for the randomly nucleated type-C  bands and model acoustic energy $R_{AE}$ plot for $\dot\epsilon_a = 1.125\times 10^{-5}/s$. (b) Expanded portion of the AE spectrum shown between the arrows in (a).}
\label{PLCAE_BAND_C}
\end{figure}

\subsection{Acoustic emission accompanying the three types of PLC bands}

The calculated plastic strain  ${\dot\epsilon}_p(k,t')$ for the entire  time interval has been used as a source term in Eqs.  (\ref{waveqn_discret1}-\ref{waveqn_discret99}) to obtain the model acoustic energy spectrum $R_{AE}$.  A plot of $R_{AE}$ along with the stress-time curve  accompanying the type-C bands  are shown in Fig. \ref{PLCAE_BAND_C} for $ {\dot \epsilon}_a =1.125\times 10^{-5}/s$.   As can be seen, the bursts of acoustic emission appear at each stress drop and are well separated  for strain rates  $3 \times 10^{-6}/s < \dot \epsilon_a < 1.5 \times 10^{-5}/s$ corresponding to the type-C bands. The post burst AE continuously decreases till a new AE burst appears.  However, on an expanded scale, we find that  there is  a build-up of the AE signal from a low level just beyond the stress drop as shown in Fig. \ref{PLCAE_BAND_C}(b). These features are confirmed by experimental AE spectra for the type-C bands  \cite{Chmelik02,Chmelik07}. 

\begin{figure}[!h]
\vbox{
\includegraphics[height=6.5cm,width=8.4cm]{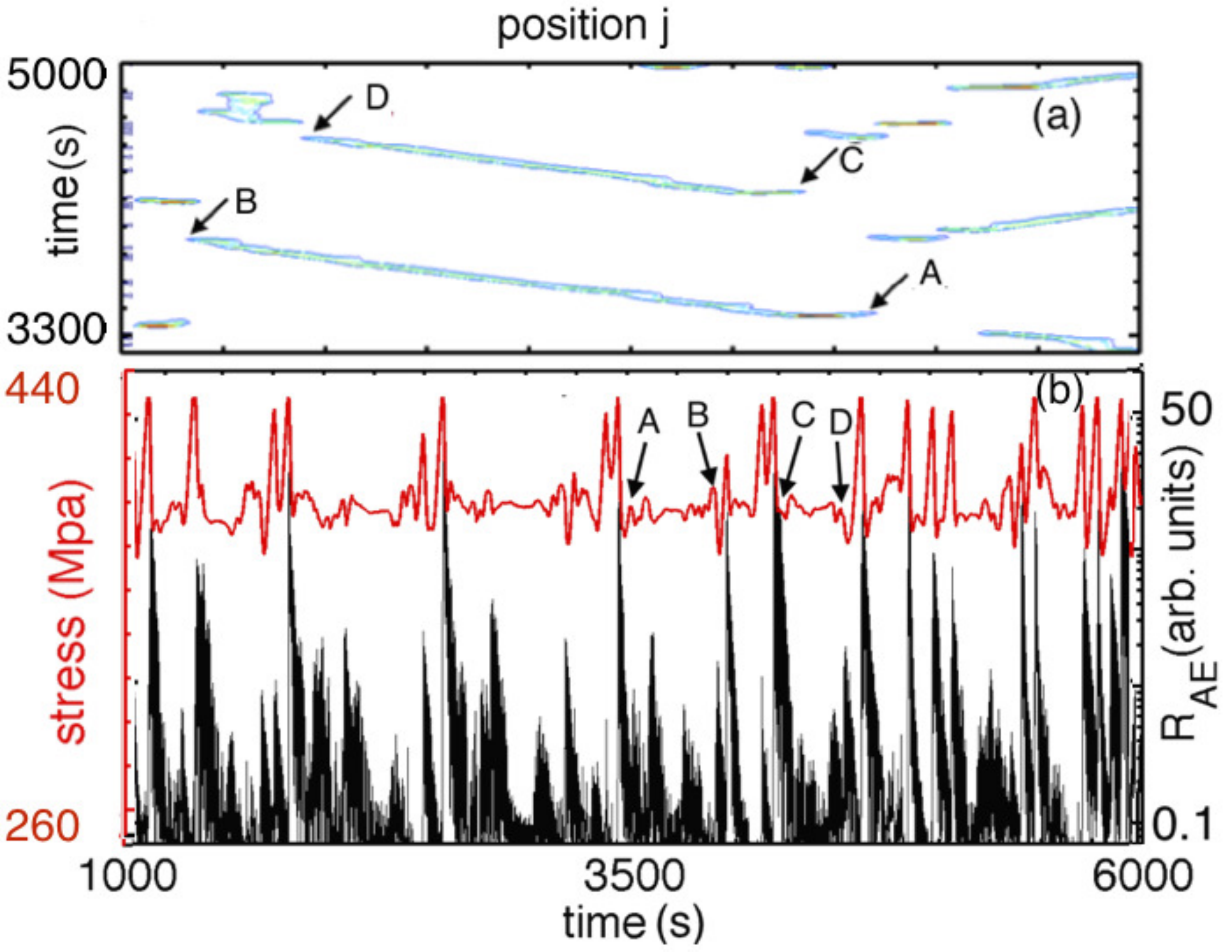}}
\caption{(color online) (a) Two partially propagating type-B bands for $\dot\epsilon_a = 3\times 10^{-5}/s$. (b) The corresponding stress-strain curve  and the model acoustic energy.}
\label{PLCAE_BAND_B}
\end{figure}

With  increasing $\dot\epsilon_a$,  the AE bursts begin to overlap.  In the region of  partially propagating  type-B bands, the AE spectrum consists of overlapping bursts leading to low amplitude continuous background.  A typical plot of the AE  spectrum for $ {\dot \epsilon}_a =3\times 10^{-5}/s$  is  shown in Fig. \ref{PLCAE_BAND_B}(b).  However,  a few large amplitude AE signals can be seen overriding the continuous low amplitude AE signals.   Two observations can be made from the figure. First, the low amplitude continuous AE signals are seen to be well correlated with the small amplitude stress serrations induced by propagating bands. Second,  the large amplitude AE bursts are well correlated with the large amplitude stress drops that  are identified with the nucleation of new bands. Thus, large bursts of AE signal are correlated with the nucleation of  new bands. This prediction is confirmed by recent experimental studies on AE during the PLC bands \cite{Chmelik02,Chmelik07}. 
\begin{figure}[!h]
\vbox{
\includegraphics[height=6.5cm,width=8.4cm]{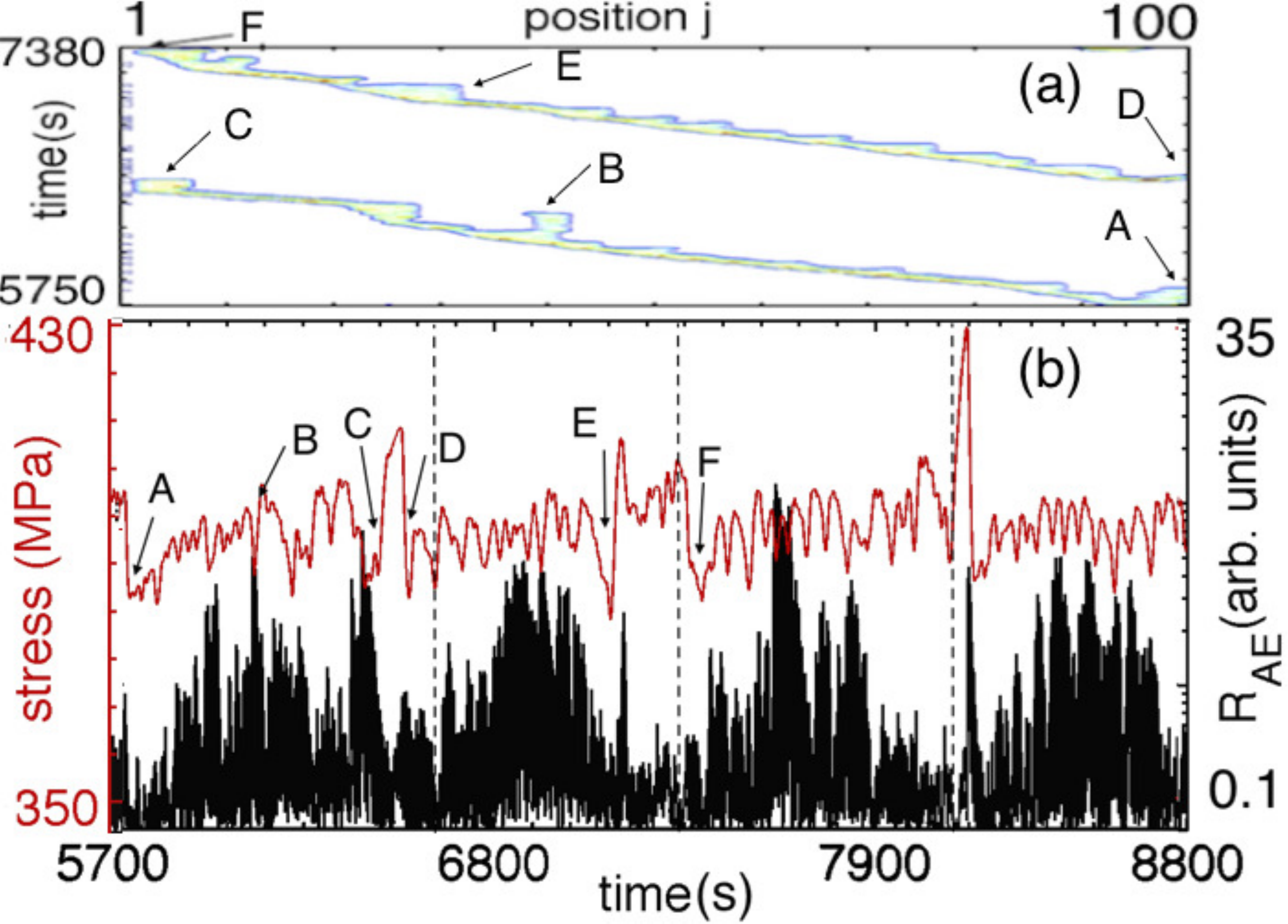}}
\caption{(color online) (a) Plots of two fully propagating type-A bands for $\dot\epsilon_a = 5.5\times 10^{-5}/s$. (b) The corresponding stress-strain curve and  the associated model acoustic energy $R_{AE}$.  }
\label{PLCAE_BAND_A}
\end{figure}

At high strain rates of the fully propagating type-A bands, the nature of the AE spectrum becomes  fully continuous.  A typical AE spectrum along with the stress-time curve for $\dot{\epsilon}_a= 5.5\times 10^{-5}/s$ is shown in Fig. \ref{PLCAE_BAND_A}(b). The figure shows that the AE spectrum is largely continuous,  a feature that is consistent with experimental AE spectrum \cite{Zeides90,Chmelik02,Chmelik07}. Two other features are also evident. The AE spectrum  exhibits occasional relatively large amplitude AE signals overriding the continuous triangular shaped AE spectrum. The relatively large burst of AE can be identified with the large amplitude stress drops at points B and E on the stress-strain curve. We also see a correlation between these large bursts of AE  with nucleation of the bands (A and D on the stress-time curve) and,  C and F when the band reaches the sample end. This identification is similar to the type-B band nucleation.  The overall triangular shape of the continuous AE signals (shown between the vertical dashed lines) is well correlated with the duration of increasing mean stress level of the SAS.   It would be interesting to verify these correlations between band propagation induced small amplitude serrations in the type-A band regime and the continuous AE spectrum.

The above discussion also shows that our approach predicts   most features of AE spectrum seen in experiments. It also   provides insight into the origin of low amplitude continuous AE spectrum seen in both the type-B and A band regimes, namely, that it is  directly correlated to the SAS's  induced by  propagating bands.  However, features associated with hardening such as the decreasing activity with strain hardening  seen experiments cannot be predicted within scope of the AK model since AK model does not include the forest hardening term. 

\section{Acoustic emission during L{\"u}ders-like propagating band }

Before proceeding further,  we briefly summarize the salient points about L{\"u}ders and show that the AK model predicts  L{\"u}ders-like  bands. L{\"u}ders bands are traditionally referred to the propagating bands seen in  polycrystalline samples  following a yield drop. The bands propagate from one end to the other at practically  zero hardening rate.  The band propagation is ascribed to  the  incompatibility stresses between the grains.  However,  L{\"u}ders-like  bands have been reported in many systems such as  single crystals, alloys doped with solutes, irradiated  single crystals and even  whiskers. (See  Ref. \cite{Neuhausser83} for a review.)  According to Neuh$\ddot a$user \cite{Neuhausser83}, the resistance offered by obstacles to dislocation motion is a common mechanism. Equivalent role of obstacles is played by  solute atoms in the AK model.  

\begin{figure}[!h]
\vbox{
\includegraphics[height=3.7cm,width=8.0cm]{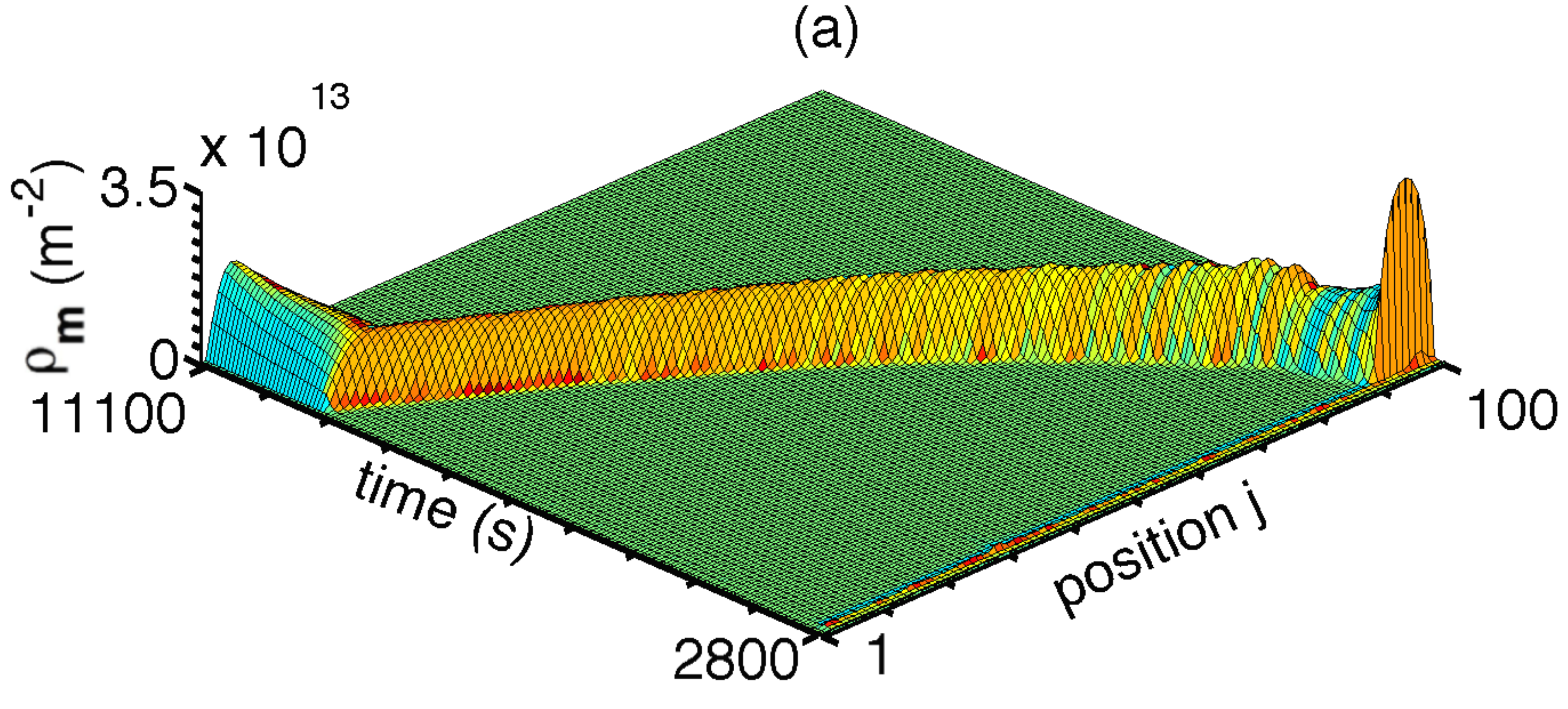}
\includegraphics[height=3.7cm,width=7.5cm]{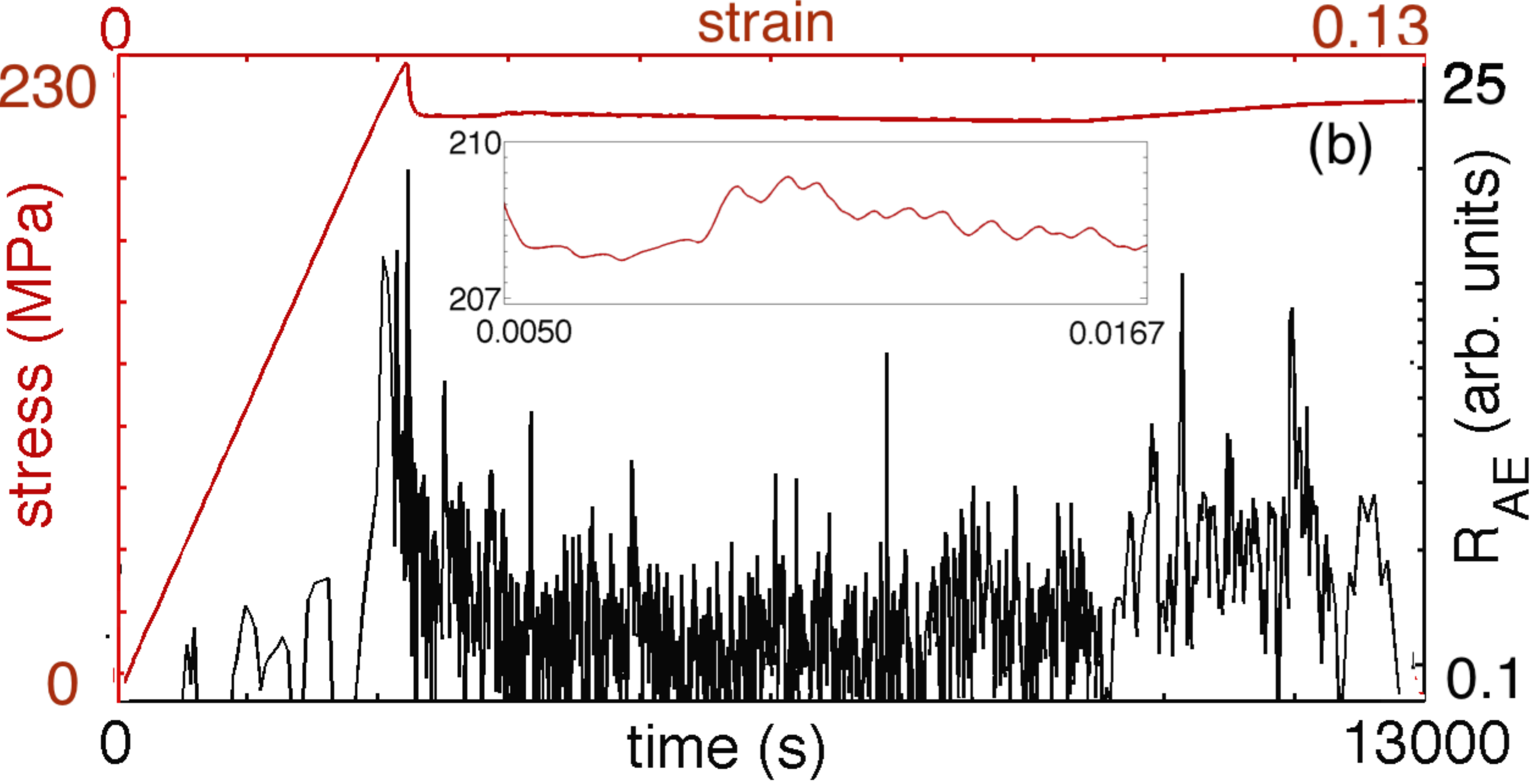}
}
\caption{(Color online) (a)  A plot of L{\"u}ders-like propagating band for $\dot\epsilon_a=1.67 \times 10^{-6}/s$. (b) The corresponding stress-strain plot.  The inset shows small amplitude stress-serrations not visible on the full scale of (b). The  model acoustic energy $R_{AE}$ is also shown. }
\label{Luders40}
\end{figure}

It is known that  alloys exhibiting the PLC effect often also exhibit L{\"u}ders regime \cite{Chmelik02,Chmelik07,Leby12}. Since the AK model exhibits most features of the PLC effect including the three band types,  it is conceivable that the AK model also exhibits  L{\"u}ders-like bands. Indeed, the AK model has recently  shown to exhibit  L{\"u}ders-like bands \cite{Ritupan15}.  Figure \ref{Luders40}(a) shows a L{\"u}ders-like  band  starting immediately after yield drop and traveling from one end to other with a near constant velocity. The  parameter values  used are $\alpha_m=1/s,\alpha_c=0.002/s, \gamma= 5\times 10^{-4}/s, E^*/\sigma_y=240,m=10$ and $ \dot\epsilon_a=1.67 \times 10^{-6}/s$.  The corresponding stress-strain curve is shown in Fig. \ref{Luders40}(b). It is clear that the stress level remains nearly constant at the  lower yield value of $\sim 200MPa$. While the stress-strain curve looks smooth on this scale, we do find small amplitude serrations as shown in the inset. 

\subsection{Acoustic emission during L{\"u}ders like band propagation}

We have calculated the AE spectrum by following the same procedure as for the previous  cases. This   is shown in Fig. \ref{Luders40}(b). As can be seen, the AE spectrum exhibits a peak corresponding to the yield drop. The AE activity rapidly  decreases to low level in the band propagation  regime. Indeed, the decrease in the peak level AE activity  in the propagating region is  consistent with experiments \cite{Chmelik02,Chmelik07,Zuev08}. The peak in the  AE spectrum at the yield is due to rapid multiplication of dislocations from a low initial density. The decrease in the AE activity during the band propagation can be identified with  band propagation induced small amplitude serrations shown in the inset of Fig. \ref{Luders40}(b).  Recall that the small amplitude serrations induced during type-A band propagation were shown to be well correlated with low amplitude continuous AE signals. In this case, the amplitude of the serrations are even smaller that the case of type-A PLC band, typically less than 2 MPa, which is  the primary  cause of the low level AE signals in propagating region.  

\section{Discussion and conclusions}

In summary, we have developed a theoretical framework for describing the nature of the AE spectrum accompanying any plastic deformation and illustrated the applicability to three distinct cases of plastic deformation. The dissipated AE energy  is represented by the Rayleigh dissipation function \cite{Rumiepl,Rajeevprl,Kalaprl,Rumiprl,Jag08}. The plastic strain rate computed from the dislocation density evolution equations  goes as a source term in the  wave equation. The several orders of magnitude difference between the  inertial time scale and the plastic deformation time scale is incorporated as a scale factor in the numerical solution of the wave equation. The necessity to impose mutually compatible boundary conditions between the wave equation and the dislocation  density evolution equations   forced us to deal with the discrete form of wave equation. 

The basic framework itself is independent of what kind  of plastic deformation is targeted as long as the plastic strain rate as a function of space and time can be calculated using some method.  However, the ability to construct the evolution equations for the dislocation microstructure matching the  experimental features  directly influences the resulting AE spectrum.  It is therefore  important to ensure that the model equations correctly  predict the observed  stress-strain curves and spatio-temporal features of the plastic deformation. Indeed, the good match between the model AE spectra and the experimental AE spectra in the three cases considered can only be attributed to the correctness in modeling the salient features of plastic deformation. The fact that the dislocation density based model closely reproduces the smooth experimental stress-strain curve as also the shape of experimental AE spectrum should be taken as a validation of the correctness of the model.  On the other hand,  the  burst like character of the AE signals reflecting  the fundamentally intermittent nature of plastic deformation  is  {\it more a validation of the correctness of the framework itself.} In the case of the PLC  effect,  the fact that the predicted AE spectra are consistent with the experimental AE spectra for each of the band types \cite{Caceres87,Zeides90,Chmelik02,Chmelik07,Leby12} is clearly  due to fact that the AK model predicts the three band types and associated serrations.  The results show that the AE spectrum consists of well separated bursts of AE occurring at every stress drop for the type-C bands. As $\dot\epsilon_a$ is increased towards the region of type-B bands,  these  bursts of AE  tend to overlap forming a low level nearly continuous background. The AE spectrum corresponding to the type-A band is nearly continuous.   The nature of the AE spectrum during L{\"u}ders-like band propagation predicted by the AK model is also consistent with experiments  \cite{Chmelik02,Chmelik07,Zuev08}. More importantly,  our method is able to capture  the intermittent burst-like character of the AE signals in all the cases considered.

Interestingly, our approach is able to predict some details  seen in experimental AE spectrum such as the unambiguous  identification of a few large amplitude AE signals  with the nucleation of a new band \cite{Chmelik02,Chmelik07}. More importantly, our study  shows that the low amplitude continuous AE spectrum seen in both the type-B and A band regimes as also  L{\"u}ders band, are directly correlated with the small amplitude serrations induced by propagating bands. Simultaneous measurement of band propagation, stress  and the associated AE spectrum should validate this result.

We now comment on  the strengths and limitations of our approach. As stated above the approach is general enough  provided the plastic strain rate can be obtained from any model that captures   the sptio-temporal features and stress-strain curve  of the specific  plastic deformation. From this point of view,  plastic strain rate obtained from H{\"a}hner \cite{Hahner02} and Kok {\it et al} \cite{Kok03} can be used to obtain the AE spectrum since these models predict the three PLC bands.  The approach can also be  applied  to other types of plastic deformations not considered here.  For example, the AE technique  has been used to study variety of modes of plastic deformation such as  load-rate controlled PLC experiments \cite{Chmelik07a}, cyclic loading experiments \cite{Chmelik04} and stress relaxation experiments \cite{Chmelik07}.  Our method is applicable to these cases since the dislocation evolution equations can be coupled to load-rate controlled and cyclic loading conditions instead of  the constant strain rate deformation \cite{Glazov97}.  The method can also be used for the case of AE studies in nano and microindentation experiments \cite{Tymiak03} since  a dislocation dynamical  model for nanoindentation has been developed recently \cite{Anan14}.

Lastly, consider the  importance of modeling dislocation evolution equations.  In alloys exhibiting the PLC effect, serrations are seen on stress-strain curves that display strain hardening. Experiments show that the  AE activity decreases with increasing strain. This feature can not be predicted by the present form of the AK model since it includes hardening only in a marginal way.  However, the strain hardening feature  is recovered once  the forest hardening term $\rho_m\rho_{im}^{1/2}$ is included in the  AK model \cite{Ritupan12a}. Therefore,  we expect that the revised AK model should recover the AE features that depend on strain hardening.  Work in this direction is in progress.

One last example is the case of acoustic emission during crack propagation \cite{Ravi04,Fruend98}. During crack propagation in ductile materials,  plastic deformation blunts the crack tip. Clearly, it should be possible to adopt our framework by using equations that describe dynamic propagation of a crack. Even in the case of brittle fracture propagation, it should be possible to adopt our  method since brittle fracture can be viewed as a limiting case of ductility.  

A comment is in order about the algorithm  followed in computing the AE spectrum.  The acoustic energy spectrum was calculated using the  plastic strain rate computed from  dislocation based models along with the machine equation Eq. (\ref{S-eqn}).  However,  Eq. (\ref{S-eqn})  assumes stress equilibration.  This was done for the sake of convenience of computation. The  framework  itself is more general since  the elastic strain can be obtained from the wave equation which can be used to obtain the unrelaxed stress. Work in this direction is in progress. 

While our approach to acoustic emission is phenomenological, recently phase field crystal (PFC) model \cite{Chan10} has been used to predict the  power law distribution of AE signals.  Since the model  has the ability to deal simultaneously with elastic and defect degrees of freedom, it  may have potential for AE studies. However,  the ability of the PFC model to predict the generic features of specific cases of plastic deformation  such as the stress-strain curve  and  the associated spatio-temporal features (such as the three types of PLC bands and L{\"u}ders band or even smooth stress-strain curve)  remains to be established since the characteristic features of AE spectrum are  directly correlated with these features. 

\section{ACKNOWLEDGMENTS}
GA acknowledges the Board of Research in Nuclear Sciences  Grant No. $2012/36/18-BRNS$ and the support from Indian National Science Academy through Senior scientist position. Part of this work was carried out when JK and RS were at the Indian Institute of Science and both were supported under the BRNS grant.  

\end{document}